\def\year{2021}\relax
\documentclass[letterpaper]{article} % DO NOT CHANGE THIS
\usepackage{aaai21}  % DO NOT CHANGE THIS
\usepackage{times}  % DO NOT CHANGE THIS
\usepackage{helvet} % DO NOT CHANGE THIS
\usepackage{courier}  % DO NOT CHANGE THIS
\usepackage[hyphens]{url}  % DO NOT CHANGE THIS
\usepackage{graphicx} % DO NOT CHANGE THIS
\usepackage{natbib}  % DO NOT CHANGE THIS AND DO NOT ADD ANY OPTIONS TO IT
\usepackage{caption} % DO NOT CHANGE THIS AND DO NOT ADD ANY OPTIONS TO IT
\usepackage{graphicx,amsmath,booktabs,latexsym}
\usepackage{amssymb}
\usepackage[lined,boxed,ruled]{algorithm2e}
\usepackage{multirow}
\usepackage{subfigure}
\usepackage[switch]{lineno}
\usepackage{hhline}

\graphicspath{{./FIG/}}

\title{Event-Triggered Communication Network with Limited-Bandwidth Constraint for Multi-Agent Reinforcement Learning}
\author{
    Guangzheng~Hu\textsuperscript{\rm 1, \rm 2},
    Yuanheng~Zhu\textsuperscript{\rm 1, \rm 2},
    Dongbin~Zhao\textsuperscript{\rm 1, \rm 2},
    Mengchen~Zhao\textsuperscript{\rm 3},
    Jianye~Hao\textsuperscript{\rm 3},
}
\affiliations{
    \textsuperscript{\rm 1} State Key Laboratory of Management and Control for Complex Systems,
    Institute of Automation \\ Chinese Academy of Sciences \\ Beijing 100190, China\\
    \textsuperscript{\rm 2} University of Chinese Academy of Sciences \\ Beijing, 100049 China\\
    \textsuperscript{\rm 3} Huawei Noah's Ark Lab \\ Beijing, China\\
    \{huguangzheng2019, yuanheng.zhu, dongbin.zhao\}@ia.ac.cn, \{zhaomengchen, haojianye\}@huawei.com
}
\begin{document}

\maketitle

\begin{abstract}
Communicating with each other in a distributed manner and behaving as a group are essential in multi-agent reinforcement learning. However, real-world multi-agent systems suffer from restrictions on limited-bandwidth communication. If the bandwidth is fully occupied, some agents are not able to send messages promptly to others, causing decision delay and impairing cooperative effects. Recent related work has started to address the problem but still fails in maximally reducing the consumption of communication resources. In this paper, we propose Event-Triggered Communication Network (ETCNet) to enhance the communication efficiency in multi-agent systems by sending messages only when necessary. According to the information theory, the limited bandwidth is translated to the penalty threshold of an event-triggered strategy, which determines whether an agent at each step sends a message or not. Then the design of the event-triggered strategy is formulated as a constrained Markov decision problem, and reinforcement learning finds the best communication protocol that satisfies the limited bandwidth constraint. Experiments on typical multi-agent tasks demonstrate that ETCNet outperforms other methods in terms of the reduction of bandwidth occupancy and still preserves the cooperative performance of multi-agent systems at the most.
\end{abstract}

\section{Introduction}\label{sec:introduction}

Deep Reinforcement Learning (DRL) has been playing a significant role and achieving remarkable success in a variety of challenging problems, such as chess games \cite{alphagozero}, real-time video games \cite{alphastar, shao2018starcraft}, robotics control \cite{levine2016end}, and image classification \cite{zhao2016vehicle}. As an extension, Multi-Agent Reinforcement Learning (MARL) has also received more and more attention in many scenarios where a stand-alone agent fails in accomplishing complicated tasks due to the lack of cooperation. The existence of multiple agents poses some common issues, such as non-static environment \cite{hernandez2017survey}, partially observability \cite{maddpg, vdn}, dimension explosion \cite{fql}, credit assignment \cite{coma, qmix}, and so on. In recent research \cite{szer2004improving}, it has been demonstrated that through internal communication, agents are able to share local information and pursue the same goal, which is important to address the nonstationarity and partially observability in multi-agent environment. Of particular interest is the distinction between two lines of research, that is hand-crafted communication protocols \cite{zhang2013coordinating} and learnable communication protocols \cite{dial, commnet}. Especially the advent of MARL allows to learn the protocols in an end-to-end way. Unfortunately, communication networks in the real world have the limited bandwidth. If there are a large number of agents and they send messages excessively, the network can be easily blocked, delaying message transmission and impairing cooperative effects.

Some research in MARL has been proposed to learn communication with limited bandwidth  \cite{atoc, schednet, pruning, imac}. But some key problems are still not well studied. Existing methods focus more on the reduction of sending behaviours but pay less attention on the explicit definition of the network bandwidth. Hence their conditions under which agents decide whether to send messages are not directly designed to fulfill bandwidth limitation.

Motivated by that, this paper proposes a new Event-Triggered Communication Network (ETCNet) to realize efficient communication in MARL faced with limited bandwidth. First of all, the limited bandwidth is translated into a penalty threshold mathematically, which is further put into an optimization problem as constraints. Then the event-triggered concept is realized in the architecture, and each sending behaviour is determined by an event-triggered module with a gating policy. The open of the gating policy poses a penalty for the occupation of bandwidth, but it enhances multi-agent cooperation because of sharing information. Therefore the synthesis of gating policy is put into a constrained Markov decision process (MDP) optimization, with the multi-agent performance as the objective and the limited bandwidth as the constraint. After introducing the Lagrange multiplier, reinforcement learning adaptively finds the optimal solution in a trial-and-error manner.

To verify the effectiveness, two typical multi-agent tasks, including cooperative navigation and predator-prey, are simulated.
We compare our method with other MARL methods that also consider the limited-bandwidth constraint, including SchedNet \cite{schednet}, Gated-ACML \cite{pruning}, and Message-dropout \cite{messagedropout}. After comparison, our ETCNet is significant in reducing the bandwidth consumption and preserves the whole system with marginal impact.

\section{Related Work}\label{sec:relatedWork}

Learning communication protocols of multi-agent systems has attracted considerable attention in literature \cite{accnet, bicnet}. Existing research directions include the message content \cite{a3c2}, the robustness of communication \cite{noisy}, metrics of emergent communication \cite{measure}, the attention mechanism of learning compacted messages \cite{peng2018attention, geng2019attention, daacmp}, memory-driven communication \cite{memorydriven}, and parameter sharing \cite{parashare}. But they pay little attention to the restriction of limited bandwidth in communication network.

To efficiently utilize finite communication resources, some recent MARL methods make agents learn to choose what, when and with whom to communicate. IC3Net \cite{ic3net} extends the work of CommNet \cite{commnet} by means of Long Short-Term Memory (LSTM) and the gating mechanism.
Gated-ACML \cite{pruning} and ATOC \cite{atoc} both evaluate the importance of communication by comparing the Q-difference between sending messages and not. If the difference is
greater than a threshold, agents consider the message is valuable and choose to communicate.
SchedNet \cite{schednet} leverages weight generators to choose top-$k$ agents with apparently more valuable observations to participate in the communication group, and broadcasts their messages to the others.
The purpose of the above methods is to reduce the bandwidth consumption but there is no mathematical definition of bandwidth constraints.
IMAC \cite{imac} argues that explicit mathematical relations exist between the entropy of messages and the bandwidth, and introduces the mutual information to approximate message entropy. By restricting the mutual information to an upper bound, the problem becomes a constrained optimization that aims to learn the efficient message generators. However, the system still transmits messages at each moment, causing the waste of communication resources if the messages at consecutive moments have similar or even the totally same content.

Another drawback of above mentioned work is that agents decide whether to communicate is just based on the current observation.
Event-triggered control is an important concept in the field of control theory to reduce the update of control signals in networked control systems \cite{et2012, zhu2016event, zhang2017event}. The occupation of communication network to send signals is conditioned on the difference of a predefined energy function between the current observation and a previous one, so the bandwidth usage is restricted.
We extend the event-triggered concept to multi-agent communication and learn communication protocols to decide when sending messages.

\section{Preliminary on DEC-POMDP with Communication}\label{sec:Background}

We consider MARL in the framework of Decentralized Partially Observable Decision Process (DEC-POMDP), which is described as a tuple $\left\langle \mathcal{S}, \boldsymbol{\mathcal{A}},P,\boldsymbol{R}, \boldsymbol{\mathcal{O}}, Z, N,\gamma \right\rangle$, where $N$ is the number of agents; $\mathcal{S}$ denotes the state space of the problem;
$\boldsymbol{\mathcal{O}} = \left\{ \mathcal{O}_i \right\}_{i=1,2,...,N}$ represents the sets of observations for each agent;
$\boldsymbol{\mathcal{A}} = \left\{ \mathcal{A}_i \right\}_{i=1,2,...,N}$ denotes the sets of actions.
$Z(s,i): \mathcal{S} \rightarrow O_i$ is the observation function that determines the private observation, and
the agent $i$ receives a private observation by $o_i = Z(s,i)$.
$P(s'|s, \boldsymbol{a}) : \mathcal{S} \times \boldsymbol{\mathcal{A}} \times \mathcal{S} \rightarrow [0,1]$ represents the state transition function, where $\boldsymbol{a}=[a_1,a_2,...,a_N]$ is the joint action.
$\boldsymbol{R} = \left\{ \mathcal{R}_i \right\}_{i=1,2,...,N}: \mathcal{S} \times \boldsymbol{\mathcal{A}} \rightarrow \mathbb{R}^N$ represents the set of reward functions.
$\gamma \in [0,1]$ denotes the discount factor.
Each agent aims to learn a policy $\pi_i(a_i|o_i): \mathcal{O}_i \rightarrow \mathcal{A}_i$ that maximizes the expected discounted return $\mathbb{E}\left[ \sum_{t=0}^\infty \gamma^t r_{i,t} \right], r_{i,t} \sim \mathcal{R}_i(s_t,\boldsymbol{a}_t)$.

Sharing observations improves the performance of the whole multi-agent system, and makes each agent learn the better policy. In such case, the policy is written by $\pi_i(a_i|o_i, \boldsymbol{o}_{-i}): \mathcal{O}_i \times \boldsymbol{\mathcal{O}}_{-i} \rightarrow \mathcal{A}_i$, where $\boldsymbol{o}_{-i} = \left[ o_{1}, ... o_{i-1}, o_{i+1}, ... o_{N}  \right] \in \boldsymbol{\mathcal{O}}_{-i}$ is the joint observation of other agents except $i$. If observations are high-dimensional, they have to be encoded to low-dimensional representations to reduce data transmission.
The policy with communication is denoted by $\pi_i(a_i|o_i, \boldsymbol{m}_{-i}): \mathcal{O}_i \times \boldsymbol{\mathcal{M}}_{-i} \rightarrow \mathcal{A}_i$, where $\boldsymbol{m}_{-i} = \left[ m_{1}, ..., m_{i-1}, m_{i+1}, ..., m_{N}  \right] \in \boldsymbol{\mathcal{M}}_{-i}$ denotes the messages that agent $i$ receives from its teammates.

In this way, we extend DEC-POMDP to a communicative one to enhance coopration. The process of communication consists of encoding, transmission, and decoding. The encoding process maps the message to a bitstream, which is transmitted through a communication channel. Decoding is the inverse operation of encoding to recover the message.
For ease of analysis, we assume the message $m$ satisfies a certain distribution $M$ and the entropy is denoted by $H(M)$. The communication network has a bandwidth $B$.
According to Shannon's source coding theorem \cite{Shannon1948}, in order to encode the message without the risk of information loss, the average number of bits $N_b$ must satisfy $ N_b \geq H(M)$.
The maximum data rate $R_{max}$ (bits per second) \cite{Freeman2004} in a noiseless channel has $R_{max} = 2B \log_2 K$, where $K$ is the number of discrete levels in the signal.
IMAC \cite{imac} combines the above two requirements together and argues that the relationship between bandwidth and message entropy has
$2B \log_2 K = R_{max} \geq n N_b \geq n H(M)$.

\section{Event-Triggered Communication Network}\label{sec:ETCNet}

Now we formally present our ETCNet. First, we detail the architecture to show its advantage of saving bandwidth and maintaining multi-agent cooperation. Second, the limited bandwidth is converted to the penalties of sending behaviours. By adding the new constraint to the multi-agent cooperative objective, the gating policy is learned by reinforcement learning to solve a constrained optimization problem.

\subsection{Architecture}\label{sec:architecture}

\begin{figure}[!ht]
   \centering
   \begin{center}
   \scriptsize
     \includegraphics*[width=3.3in]{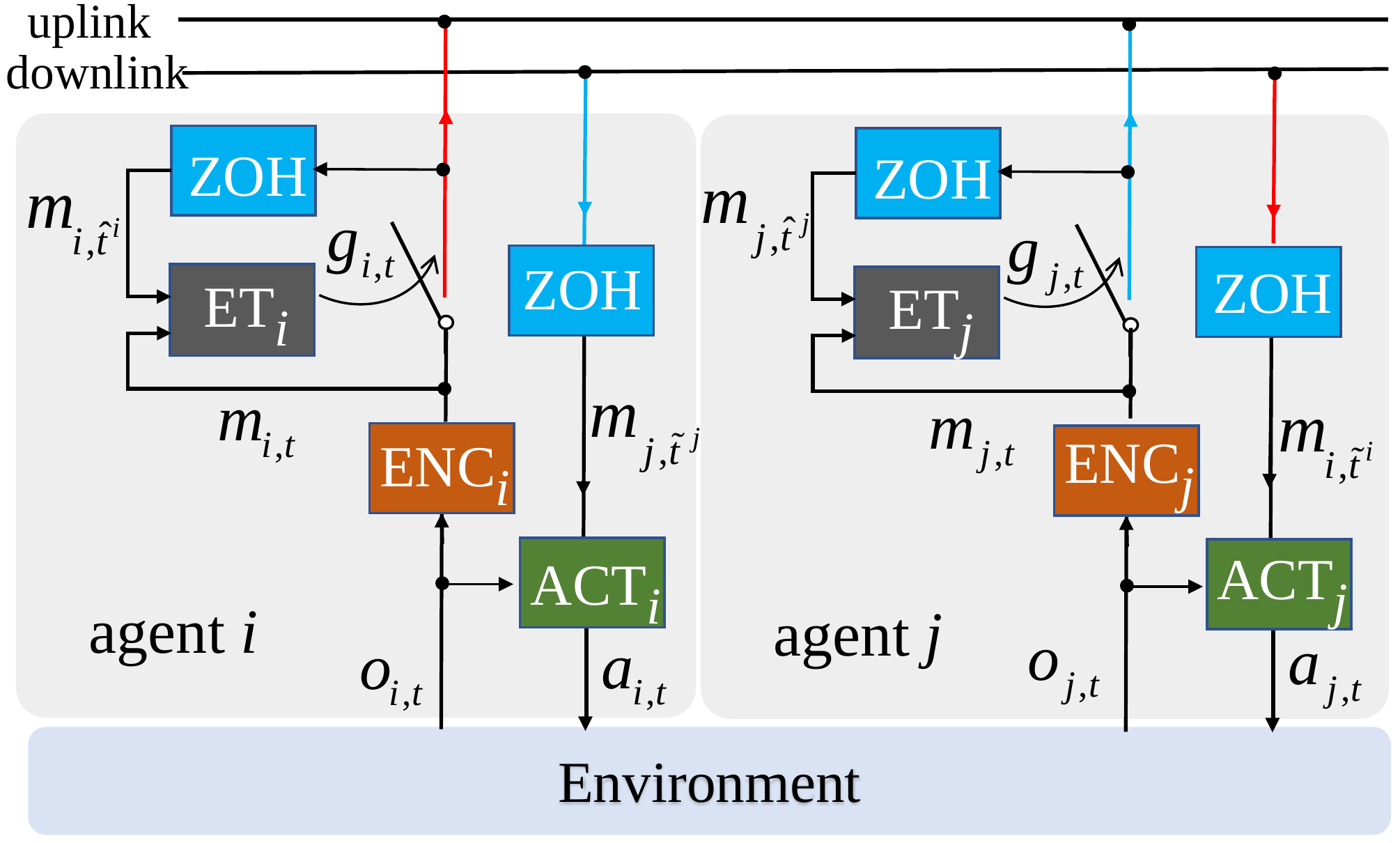}\\
   \caption{Framework of the proposed ETCNet. ENC$_i$ is the encoding module; ET$_i$ is the event-triggered gating module; ACT$_i$ is the agent policy module; ZOH represents zero-order holder.}
   \label{fig:architecture}
   \end{center}
\end{figure}

Figure \ref{fig:architecture} presents the architecture of ETCNet in multi-agent settings.
The execution process of each agent consists of three phases: ($i$) observation encoding; ($ii$) message gating and sending; ($iii$) message receiving and decision making.
The gating module is designed in an event-triggered way such that the sending behaviour happens only if it is necessary. At the message-receiving and decision-making stage, if an agent opens its transmission gate, other agents will receive its current message to decide cooperative actions. Otherwise, they will use the lastly received message, memorized by a zero-oreder holder (ZOH) module, to continue to cooperate.

Before digging into the detailed design, we list some key notations as follows: consider $N$ agents, for agent $i$ at time $t$, its observation is denoted by $o_{i,t}$; its current message is $m_{i,t}=e_{i}\left( o_{i,t} \right)$, where $e_{i}(\cdot)$ is the encoding function; its gating action is denoted by $g_{i,t} \sim \mu_{i}(\cdot)$ , where $\mu_{i}(\cdot)$ represents the gating policy function; its action executed on the environment is denoted by $a_{i,t} \sim \pi_{i}(\cdot)$ , where $\pi_{i}(\cdot)$ represents the agent policy function.
The gating action samples from $\{0,1 \}$, where $1$ represents the event is triggered and the transmission is open. Otherwise, $0$ is not.

\begin{figure}[!ht]
   \centering
   \begin{center}
   \scriptsize
     \includegraphics*[scale=0.75]{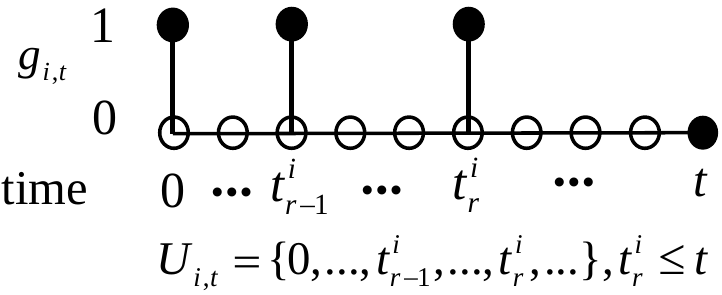}\\
   \caption{Illustration of $U_{i,t}$ whose elements are event-triggered time points of agent $i$.}
   \label{fig:U}
   \end{center}
\end{figure}

We specify $U_{i,t} = \left[ t^i_0, ..., t^i_r, ... \right]$ to denote the set of event-triggered time points $t^i_r$ at the current $t$, shown as Figure \ref{fig:U}.
The more explicit expression for the above mentioned variables and functions is presented.
The gating policy is denoted by $g_{i,t} \sim \mu_{i} ( m_{i,t}, m_{i,\hat{t}^i} )$, where $m_{i,\hat{t}^i}$ represents the message at the lastly triggering moment, memorized by ZOH, and $\hat{t}^i = \underset{\kappa \in U_{i,t-1}}{\arg \min } \left \{ t - \kappa \right \}$). Note that $U_{i,t-1}$ could not be updated to $U_{i,t}$ before agent $i$ makes gating decisions at $t$.
An agent chooses to send messages only when it considers the change of two inputs will facilitate the cooperation.
The agent action follows $a_{i,t} \sim \pi_{i}( o_{i,t}, \tilde{\boldsymbol{m}}_{-i,t})$, where $\tilde{\boldsymbol{m}}_{-i,t} = \left[ m_{1,\tilde{t}^1}, ..., m_{i-1,\tilde{t}^{i-1}}, m_{i+1,\tilde{t}^{i+1}}, ..., m_{N,\tilde{t}^N} \right]$, where $\tilde{t}^j= \underset{\kappa \in U_{j,t}}{\arg \min } \left \{ t - \kappa \right \}$.
In addition to own observations, the policy uses received messages (if there are) or memorized messages from others to realize cooperation.

Compared with existing work, the biggest difference in architecture is that our ETCNet not only uses the current observation to define the sending condition, but also relies on the lastly sent message.
Beyond that, if no message is received, the agent uses the memorized message rather than the zero vector (e.g. Gated-ACML \cite{pruning}, ATOC \cite{atoc}, and SchedNet \cite{schednet}) to prevent the lose of information and preserve the cooperation performance.

\subsection{Limited-bandwidth Constraint and Penalty Threshold}\label{sec:definition}

According to the Preliminary, the maximum symbols per second on a limited-bandwidth channel satisfy
\begin{equation} \label{eq:n1}
\begin{gathered}
n  \leq \frac{2B \log_2 K}{H(M)}.
\end{gathered}
\end{equation}
However, the distribution and entropy of message $M$ is generally unknown, and all we can get are its statistic properties like mean and variance. The principle of maximum entropy  \cite{principle} proves that the Gaussian distribution has the maximum entropy compared with other probability distributions with the same mean and variance.
We can take the entropy of a Gaussian distribution $H(X)=\log (2\pi e \sigma^2), X \sim N(\mu, \sigma)$ as an upper bound of $H(M)$, where $\mu$ and $\sigma$ are the mean and variance of $M$.
Substituting it back to (\ref{eq:n1}) yields
\begin{equation} \label{eq:n2}
\begin{gathered}
n \leq \frac{2B \log_2 K}{H(X)} = \frac{4B \log_2 K}{\log (2\pi e \sigma^2)}.
\end{gathered}
\end{equation}

Suppose the gating policy has a probability of $p$ sending messages at each step, and a message has a length of $L$ symbols. The system sampling frequency is $F$. For a number of $N$ agents, the average number of symbols on the channel is equal to $N(N-1)LFp $ and should be no greater than $n$.
Then we are able to deduce an upper bound of probability that each agent is allowed to send messages at each step
\begin{equation} \label{eq:p2}
\begin{gathered}
p \leq p_{sup} = clip \left( \frac{4B \log_2 K}{\log (2\pi e \sigma^2) N(N-1)LF}, 0, 1 \right).
\end{gathered}
\end{equation}

Since sending messages or not corresponds to the open or close of the gating policy, so we can describe the occupation of bandwidth as a sum of penalties over the time horizon:
\begin{equation} \label{eq:C}
\begin{gathered}
C = \mathbb{E}\left[ \sum_{t=0}^{\infty} \gamma^t \mathbb I(g_{i,t}=1) \right]  \leq \frac{p_{sup}}{1-\gamma}=C_{sup}
\end{gathered}
\end{equation}
where $C_{sup}$ indicates the penalty threshold, and $\mathbb I(g_{i,t}=1)$ specifies the instantaneous penalty when an agent occupies the bandwidth.

With the original sum of rewards as the optimization objective, the problem now becomes solving the constrained MDP
\begin{equation} \label{eq:optimization}
\begin{split}
\max\,\, \mathbb{E}\left[ \sum_{t=0}^\infty \gamma^t r_{i,t} \right], \quad
s.t.\quad
\mathbb{E}\left[ \sum_{t=0}^\infty \gamma^t  g_{i,t} \right] \leq C_{sup}.
\end{split}
\end{equation}

Note that IMAC \cite{imac} also gives an explicit mathematical transformation of bandwidth limitition.
It reduces bandwidth occupation through message compression, but the transmission frequency is unchanged.
Our ECNet keeps the completeness of information and reduces the frequency of sending messages in an event-triggered way.

\subsection{Optimization Algorithm}\label{sec:algorithm}

In the implementation of ETCNet, we define three networks for each agent: EncoderNet, GatingNet, and ActorNet. They correspond to the encoding function, the gating policy function, and the agent policy function, respectively. All the homogeneous agents share the same models rather than defining different network parameters.
To lower down the learning difficulty of three networks, we separate the training into two processes. First, we train the EncoderNet and ActorNet at the full communication, that is the event-triggered module always sends message. After obtaining the well-trained  EncoderNet and ActorNet, we keep them fixed and train the GatingNet.

\textbf{Training EncoderNet and ActorNet.} The Centralized Training and Decentralized Execution (CTDE) paradigm is adopted to train EncoderNet and ActorNet to overcome the non-stationary problem.
Typically we use a centralized CriticNet parameterized by $\theta_c$ to estimate the state value function $V_{\theta_c}(\upsilon_{i}) \approx \mathbb{E}\left[ \sum_{t=0}^\infty \gamma^t r_{i,t} \right]$ where $\upsilon_{i}=[o_{i},\boldsymbol{o}_{-i}]$ is taken as the approximation of the global state $s$. The EncoderNet and ActorNet are parameterized by $\theta_e$ and $\theta_a$, respectively.
The critic value is updated based on temporal-difference as
\begin{align}
\delta_{c} & = r_{i,t} + \gamma V_{\theta_c}(\upsilon_{i,t+1}) - V_{\theta_c}(\upsilon_{i,t}), \\
\mathcal{L}_{i,t}^{critic} & =  \delta_{c}^{2}.
\end{align}
The EncoderNet and ActorNet train the parameters by back-propagation of the policy loss
\begin{multline}
\mathcal{L}_{i,t}^{act\_enc} = -\log{\pi_{i}(a_{i,t}|o_{i,t},\boldsymbol{m}_{-i,t},\theta_a)}\delta_{c}  \\
- \alpha H(\pi_{i}(a_{i,t}|o_{i,t},\boldsymbol{m}_{-i,t},\theta_a))
\end{multline}
where
$\boldsymbol{m}_{-i,t} = \left[ m_{i,t}, ..., m_{i,t}, m_{i+1,t}, ..., m_{N,t} \right]$ for full communication and $m_{i,t} = e(o_{i,t} | \theta_e)$ . An entropy term is used to discourage premature convergence.

\textbf{Training GatingNet.}
After learning the EncoderNet and ActorNet at the pretraining stage, we apply them to ETCNet framework and keep their parameters fixed. Now we learn the GatingNet parameterized by $\theta_g$, for the gating policy $\mu_{i} ( m_{i,t}, m_{i,\hat{t}^i} |\theta_g )$ to satisfy the constrained optimization as (\ref{eq:optimization}).
We use a Lagrangian multiplier $\lambda \ge 0$  to deal with the constraint and define the Lagrangian function
\begin{equation} \label{eq:lagrangian}
\begin{split}
L(\mu_{i},\lambda)=\mathbb{E}\left[ \sum_{t=0}^\infty \gamma^t (r_{i,t} -\lambda g_{i,t}) \right]+\lambda C_{sup}.
\end{split}
\end{equation}
The dual objective $d(\lambda)$ of the primal problem is defined as
\begin{equation} \label{eq:d}
\begin{split}
d(\lambda) = \sup_{\mu_{i}} L(\mu_{i},\lambda).
\end{split}
\end{equation}

Suppose at step $t$, we have had an estimate of $\lambda$, denoted as $\lambda_t$. The optimal solution for (\ref{eq:d}) is to find $\mu_i^*=\arg\underset{\mu_{i}}{\max}\, L(\mu_{i},\lambda_t)$,
which is in fact reduced to solve a new MDP optimization with the new reward $r'_{i,t}=r_{i,t}-\lambda_t g_{i,t}$.
Reinforcement learning is able to optimize the multi-agent performance considering the new reward signal by updating the gating policy with the learned EncoderNet and ActorNet.
A centralized LagrangianNet parameterized by $\theta_L$ is used to estimate the state value function $V_{\theta_L}(\upsilon_{i}) \approx \mathbb{E}\left[ \sum_{t=0}^\infty \gamma^t r'_{i,t} \right]$ for the GatingNet, and the value and policy losses are defined by
\begin{align}
\delta_L & = r'_{i,t} + \gamma V_{\theta_L}(\upsilon_{i,t+1}) - V_{\theta_L}(\upsilon_{i,t}), \\
\mathcal{L}_{i,t}^{Lagr} & = \delta_L ^2, \\
\mathcal{L}_{i,t}^{gate} & = -\log{\mu_{i}(g_{i,t}|m_{i,t}, m_{i,\hat{t}^i},\theta_g)}\delta_L \nonumber \\
& \qquad \qquad \quad - \alpha H(\mu_{i}(g_{i,t}|m_{i,t}, m_{i,\hat{t}^i},\theta_L)).
\end{align}
Note that $\lambda_t$ is an estimate of the true $\lambda$ and the optimal multiplier $\lambda^*$ satisfies
\begin{equation} \label{eq:lambda_star}
\begin{split}
\lambda^* = \arg\underset{\lambda \ge 0}{\min}\, g(\lambda).
\end{split}
\end{equation}
Then the $\lambda_t$ is updated following
\begin{align}
\lambda_{t+1} & = (\lambda_{t}-\eta_{\lambda} \nabla d(\lambda))^+ , \\
\nabla d(\lambda) & =\frac{\partial d(\lambda)}{\partial \lambda}=\left.\frac{\partial L\left(\mu_{i}, \lambda\right)}{\partial \lambda}\right|_{\mu_{i}=\mu_i^*} \nonumber \\
&=-\mathbb{E}\left[\sum_{t=0}^{\infty} \gamma^{t} g_{i, t}\right]+C_{sup}.
\end{align}
A PenaltyNet parameterized by $\theta_p$ is used to estimate the penalty value function to approximate $V_{\theta_p}(\upsilon_{i}) \approx \mathbb{E}\left[ \sum_{t=0}^\infty \gamma^t g_{i,t} \right]$. Its parameters are updated based on temporal-difference
\begin{equation} \label{eq:penalty_g}
\begin{gathered}
\mathcal{L}_{i,t}^{penalty} = \left[ g_{i,t} + \gamma V_{\theta_p}(\upsilon_{i,t+1}) - V_{\theta_p}(\upsilon_{i,t}) \right]^2.
\end{gathered}
\end{equation}
Then the update of $\lambda$ becomes
\begin{equation}
\begin{split}
\lambda_{t+1} = (\lambda_{t}-\eta_{\lambda} (-V_{\theta_p}+C_{sup}))^+.
\end{split}
\end{equation}

Considering the variance of messages varies with the change of the gating policy, we calculate the variance and update the penalty threshold $C_{sup}$ periodically throughout the training. At the end of gradient iterations, the optimal policy $\mu_{i}^*$ for the unconstrained problem is obtained. The pseudocode of training ETCNet is presented in the supplementary material.
Note that in the field of control theory, event-triggered control mainly forcus on reducing the update of control signals, and its triggering condition relies on a predefined energy function \cite{et2012, zhu2016event, zhang2017event}. Here the event-triggered module is applied to reduce the transmission of messages, and the communication protocols are learned from scratch.

\section {Experiments}\label{sec:Experiment}

\begin{figure}[h]
\begin{center}
\subfigure[Cooperative Navigation]{
\includegraphics[width=0.4\linewidth]{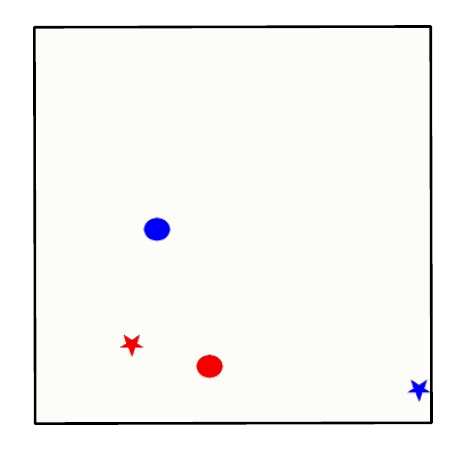}
}
\subfigure[Predator and Prey]{
\includegraphics[width=0.4\linewidth]{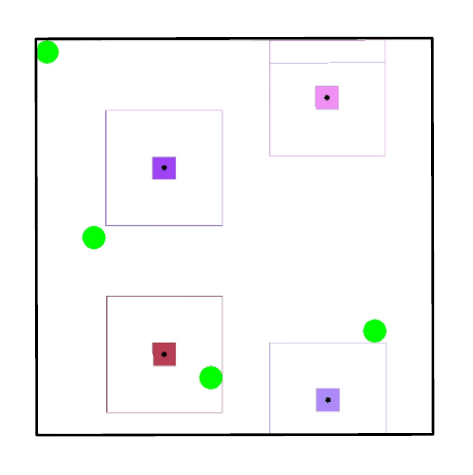}
}
\caption{Illustrations of two multi-agent tasks. (a) A circle represents an agent and the pentagram with the same color represents its destination. (b) A green circle represents a prey and a small square represents a predator with the local view surrounded by a large square. }
\label{fig:tasks}
\end{center}
\end{figure}

Two variants of multi-agent partical environments are introduced to test the performance of ETCNet on saving communication resurces, that is Cooperative Navigation and Predator and Prey \cite{maddpg} shown in Figure \ref{fig:tasks}.
The detailed experimental configurations are expatiated in the following subsections as well as in the supplementary material. \\
\textbf{Cooperative Navigation.} There are two agents and each agent aims to arrive at a specified and dynamic destination by moving along discrete directions. Each agent only observes the destination and position of the other agent. Once an agent reaches its destination, it will get a positive reward. Otherwise there is always a negative reward until the end of episode. \\
\textbf{Predator and Prey.} In this scenario, $n$ predators chase $m$ preys within a certain area. The top and the bottom, and the left and the right of the area are interconnected.
Predators and preys have the same velocity, and preys are equipped with a fixed escape policy (running from the closest predator) with a complete map vision.
A predator only has a local view around itself, so they have to cooperate to capture preys and are required to avoid collision with other predators.

Our baselines for limited bandwidth are (1) Gated-ACML \cite{pruning}, (2) SchedNet \cite{schednet}, and (3) Message-Dropout \cite{noisy}. In addition, A3C2 \cite{a3c2} is introduced as a full-communication version. In fairness, the communication protocols of all methods are set to the same as A3C2, that is, each agent sends the same encoded message to the others.
Ideally, A3C2 allows agents to fully communicate and should have the best performance. In Message-dropout, agents are allowed to send messages only with a certain probability, so it can be seen as the randomly-failed-communication version of A3C2. The gate module in Gated-ACML works much the same as the attention module in ATOC \cite{atoc}, so we choose Gated-ACML as the baseline to represent the class of $\Delta Q$-based algorithms.
Note that SchedNet selects top-$k$ agents to send messages at each step, leading to the discrete property of sending probability in the $N$-agent system, such as $1/N$ and $k/N$.
For the sake of fairness, we set identically desired sending probabilities for every baseline by adjusting some parameters, such as bandwidth of ETCNet, Q-difference threshold of Gated-ACML, and the probability of dropout in Message-Dropout.
Other configurations are given in detail in the supplementary material.

\subsection{Cooperative Navigation}\label{sec:ng}

In this task, we consider communication network with $K=2$ and $L=6$. The sampling frequency of the system has $F=45$Hz. The bandwidth is first limited to 170 bit/s. At the full communication of ETCNet, the variance of massages is $\sigma^2=0.69$, so the maximally allowed communication probability is about $50\%$. After calculating the penalty threshold and continuing the GatingNet training in the event-triggered architechture, we observe that the messsage variance varies slightly smaller to $\sigma^2=0.57$, which further relaxes the upper bound of communication probability. In fact, the final ETCNet agents send messages at each step only with 46\% probability, lower than the desired 50\%.

We take the number of steps accomplishing the task as the evaluation. The fewer steps, the better performance. Table \ref{tab:ng1} shows the results of ETCNet and baselines. It is observed that ETCNet is far superior to other methods under the same communication constraint, and is closest to the performance of full communication.
\begin{table}[h]
  \centering
  \caption{The performance of ETCNet and baselines for Cooperative Navigation with $50\%$ communication.}
    \begin{tabular}{|c|c|}
    \hline \hline
    {\bf Methods} & {\bf Steps}  \\ \hline
    \bf{ETCNet} & $\bf{16.50 \pm 6.03}$ \\
    Gated-ACML & $33.83 \pm 10.08$  \\
    SchedNet & $39.49 \pm 3.27$ \\
    Message-Dropout & $35.09 \pm 7.96$ \\ \hline
    A3C2(Full communication) & $16.20 \pm 5.85$  \\
    \hline \hline
    \end{tabular}%
  \label{tab:ng1}%
\end{table}%

\begin{figure}[h]
\begin{center}
\subfigure[Step evaluation]{
\includegraphics[width=3in, height=1.2in]{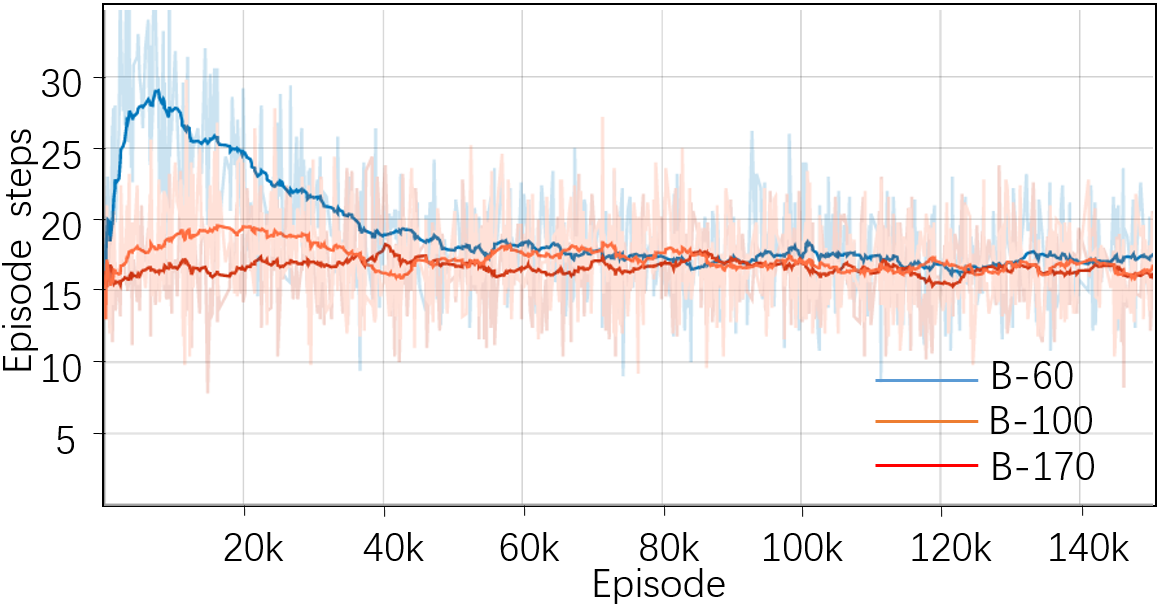}
}
\subfigure[Mean penalty per step]{
\includegraphics[width=3in, height=1.2in]{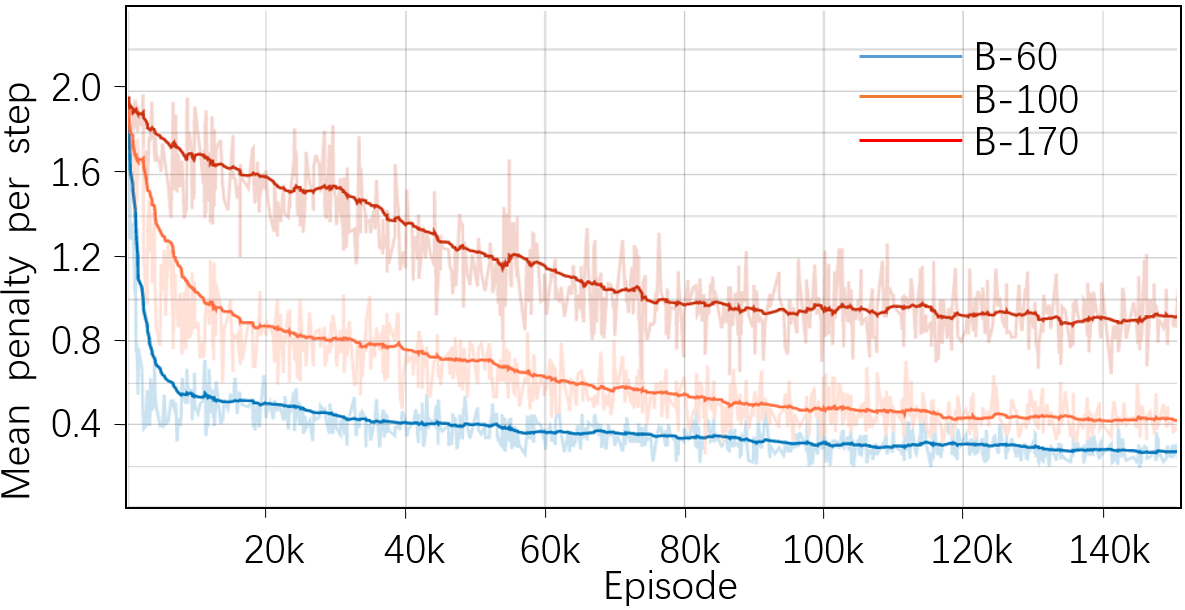}
}
\caption{Learning curves for Cooperative Navigation with respect to different bandwidths.}
\label{fig:nglearningcurves}
\end{center}
\end{figure}

To demonstrate that ETCNet can greatly reduce bandwidth consumption and preserve the multi-agent cooperation, we repeat the experiment with $B = 100$ bit/s and $B =60$ bit/s.
Figure \ref{fig:nglearningcurves} shows the learning curves of ETCNet with different bandwidths.
The top subgraph shows that the more limitation on bandwidth, the more degradation of performance in the early stage. But the curves are still stabilized back to near optimality through the later training. The bottom subgraph shows that all experiments optimize the sending penalties to satisfy corresponding thresholds. The final communication percentages are $14\%$, $24\%$ and $46\%$ for $B$ equal to 60, 100, and 170, respectively. It is concluded that ETCNet is capable of adjusting to different bandwidth constraints and preserving the best performance.

\subsection{Predator and Prey}\label{sec:pp}

In this task, we first consider 3-agent Predator and Prey. The communication networks has $K=2$ and $L=15$. The sampling frequency of the system has $F=45$Hz. The bandwidth is first limited to 580 bit/s. At the full communication of ETCNet, the variance of massages is $\sigma^2=0.330$, so the maximally allowed communication probability is about $33.3\%$. After calculating the penalty threshold and continuing the GatingNet training in the event-triggered architecture, we observe that the message variance barely changes, so the bandwidth limitation is not violated.

We take the number of steps accomplishing the task as the evaluation. The fewer steps, the better performance. Table \ref{tab:pp1} gives the performance of ETCNet and baselines with 3 agents and $33.3\%$ desired communication probability. It shows that SchedNet is competitive to our ETCNet and they both achieves the similar performance to the full-communication results.
We further compare their performance under different communication probabilities and agent numbers. The results are also listed in Table \ref{tab:pp1}. In some experiments, ETCNet outperforms all baselines, while in the others, it is competitive to the best SchedNet. The performance gap between ETCNet and the full communication is quite small. It is worth noting that ETCNet works in a variety of limited-bandwidth settings and optimally exploits the bandwidth. The communication probability in SchedNet is proportional to
$1/N$ since its mechanism is to select top-$k$ agents to send messages at each step.
\begin{table*}[t]
  \centering
  \caption{The performance of ETCNet and baselines for Predator and Prey with different agent numbers and communication upper bound.}
    \begin{tabular}{|c|c|c|c|c|c|}
    \hline \hline
      \multirow{2}{*} & {2 agents} & \multicolumn{2}{|c|}{3 agents} & \multicolumn{2}{|c|}{4 agents}  \\  \cline{2-6}
     ~ & {$\le $ 50\%} & {$\le $ 33.3\%} & {$\le $ 66.6\%} & {$\le $ 25\%} & {$\le $ 50\%}  \\ \hline
    \bf{ETCNet} & $\bf{52.19 \pm 14.63}$   & $\bf{54.59 \pm 18.63}$ & $53.52 \pm 17.81$ & $47.79 \pm 19.02$ & $\bf{46.04 \pm 17.39}$ \\
    {Gated-ACML} & $93.80 \pm 13.33$ & $65.87 \pm 22.05$ & $60.35 \pm 20.40$ & $70.67 \pm 28.86$ & $46.28 \pm 18.43$  \\
    {SchedNet} & $54.06 \pm 13.14$ & $54.95 \pm 17.41$ & $\bf{52.78 \pm 17.71}$ & $\bf{45.52 \pm 16.69}$ & $46.66 \pm 18.05$  \\
    {Message-Dropout} & $91.385 \pm 13.94$ & $85.33 \pm 19.32$ & $61.01 \pm 19.53$ & $73.933 \pm 26.404$ & $59.63 \pm 23.98$ \\ \hline
    {A3C2 (Full communication)} & {$51.08 \pm 12.68$} & \multicolumn{2}{|c|}{$50.68 \pm 16.75$} & \multicolumn{2}{|c|}{$45.256 \pm 16.755$}  \\
    \hline \hline
    \end{tabular}%
  \label{tab:pp1}%
\end{table*}%

\begin{figure}[h]
\begin{center}
\subfigure[Step evaluation]{
\includegraphics[width=2.9in, height=1.2in]{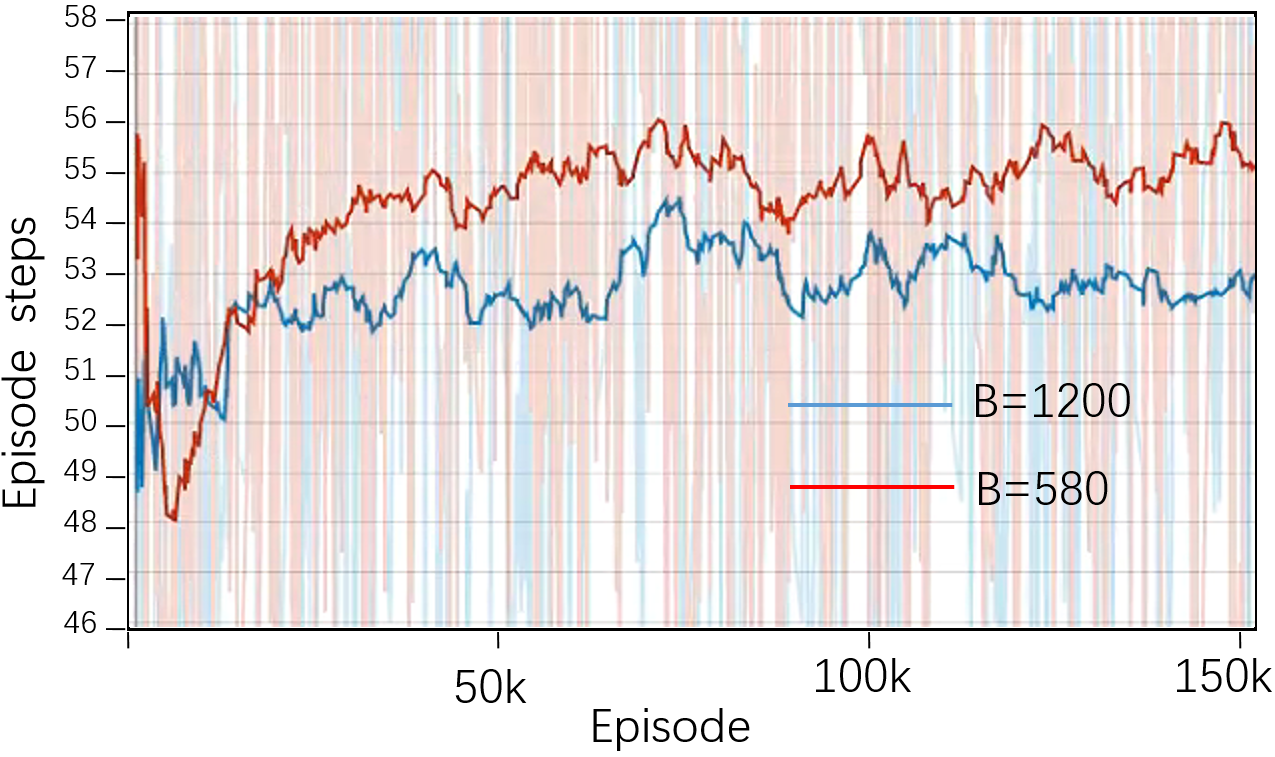}
}
\subfigure[Mean penalty per step]{
\includegraphics[width=3in, height=1.2in]{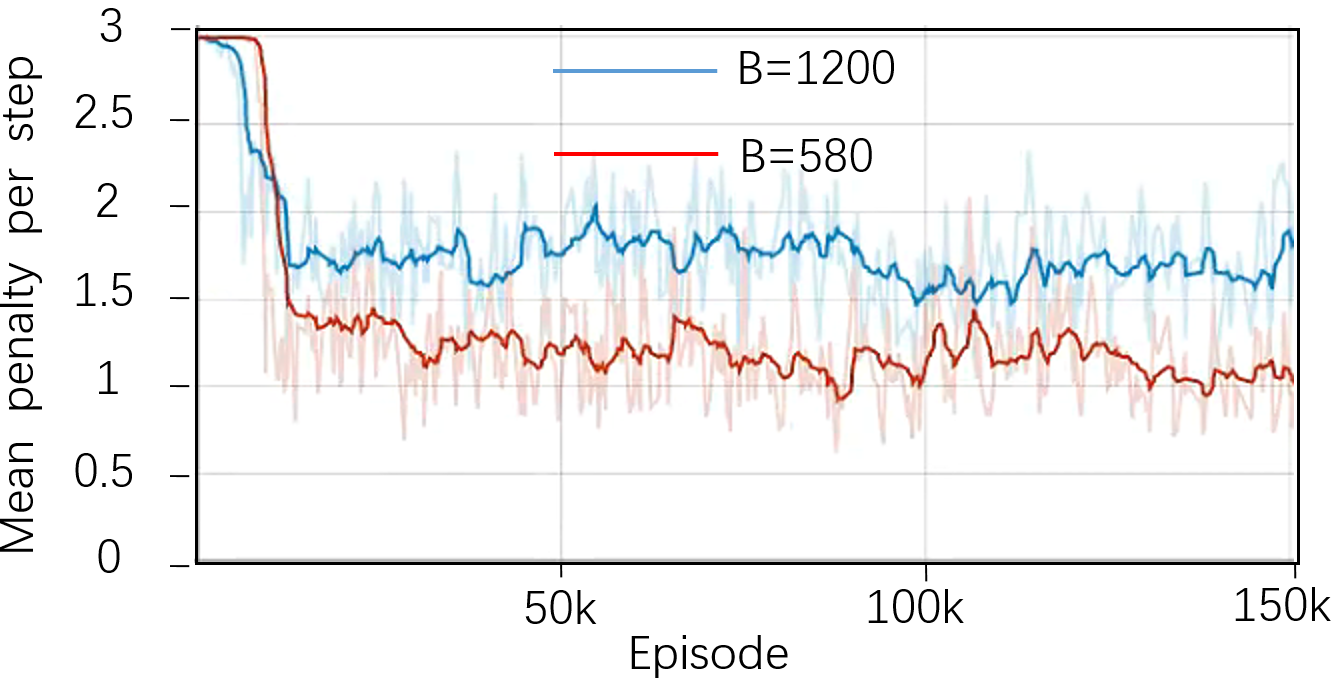}
}
\caption{Learning curves for Predator and Prey with respect to different bandwidths.}
\label{fig:pplearningcurves}
\end{center}
\end{figure}

Figure \ref{fig:pplearningcurves} shows the learning curves of ETCNet for 3-agent Predator and Prey with different bandwidths ($B = 580$ bit/s and $B = 1200$ bit/s). The learned gating policies send messages at probabilities $p=31.7\% $ and $p = 55.7\%$, respectively.
The plot shows that the learning process with the lower bandwidth has the lower frequency of sending messages, but the evaluation is worse than the learner with higher bandwidth. It is consistent with the fact that more communication is benificial to multi-agent cooperation.

\subsection{Event-triggered Gate}\label{sec:ET}
\begin{figure*}[t]
   \centering
   \begin{center}
   \scriptsize
     \includegraphics*[width=6.5in]{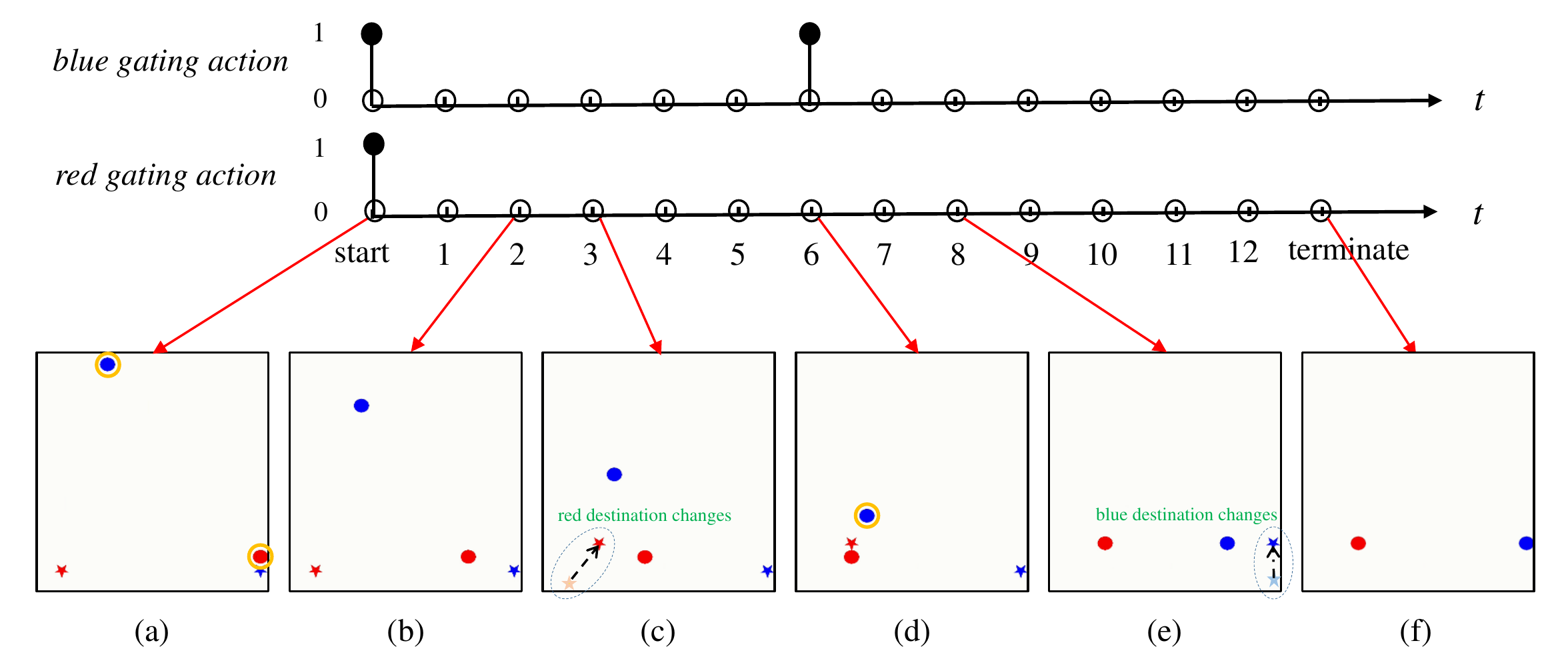}\\
   \caption{An event-triggered gating display in an epoch. The circle represents an agent and the pentagram with the same color represents its destination. The yellow ring surrounding an agent indicates it is currently sending message. }
   \label{fig:time}
   \end{center}
\end{figure*}

We argue that agents in ETCNet send messages only when necessary. We demonstrate this argument by analyzing system trajectories of Cooperative Navigation obtained by ETCNet.
Figure \ref{fig:time} shows gating actions and representative sceneries in an epoch.
We first focus on the sending behaviours of blue agent. It sends message at the starting point (a) and does not send at (b) because of no changes in its observations. It even refuses to send message at (c) when the red destination moves. It is because the lastly  received message of the red agent can still help in choosing the correct action (towards the left), especially considering that the red destination is likely to change later. It sends message at (d) because the red agent will be mislead to the wrong direction by the old message.
Now let us see the gating of red agent. It does not send message at (e) although its observation changes with the blue destination. The blue agent continues to utilize the old message and moves forwards the correct direction. When the blue agent reaches its destination at (f), the epoch terminates with both agents accomplishing their tasks. It suggests that ETCNet agents trigger
the gating policy only when the communication is important for cooperation, not simply determined by the change of observation.

\subsection{Ablation}\label{sec:Ablation}

\begin{figure}[htbp]
\begin{center}
\subfigure[Step evaluation]{
\includegraphics[width=2.94in, height=1.3in]{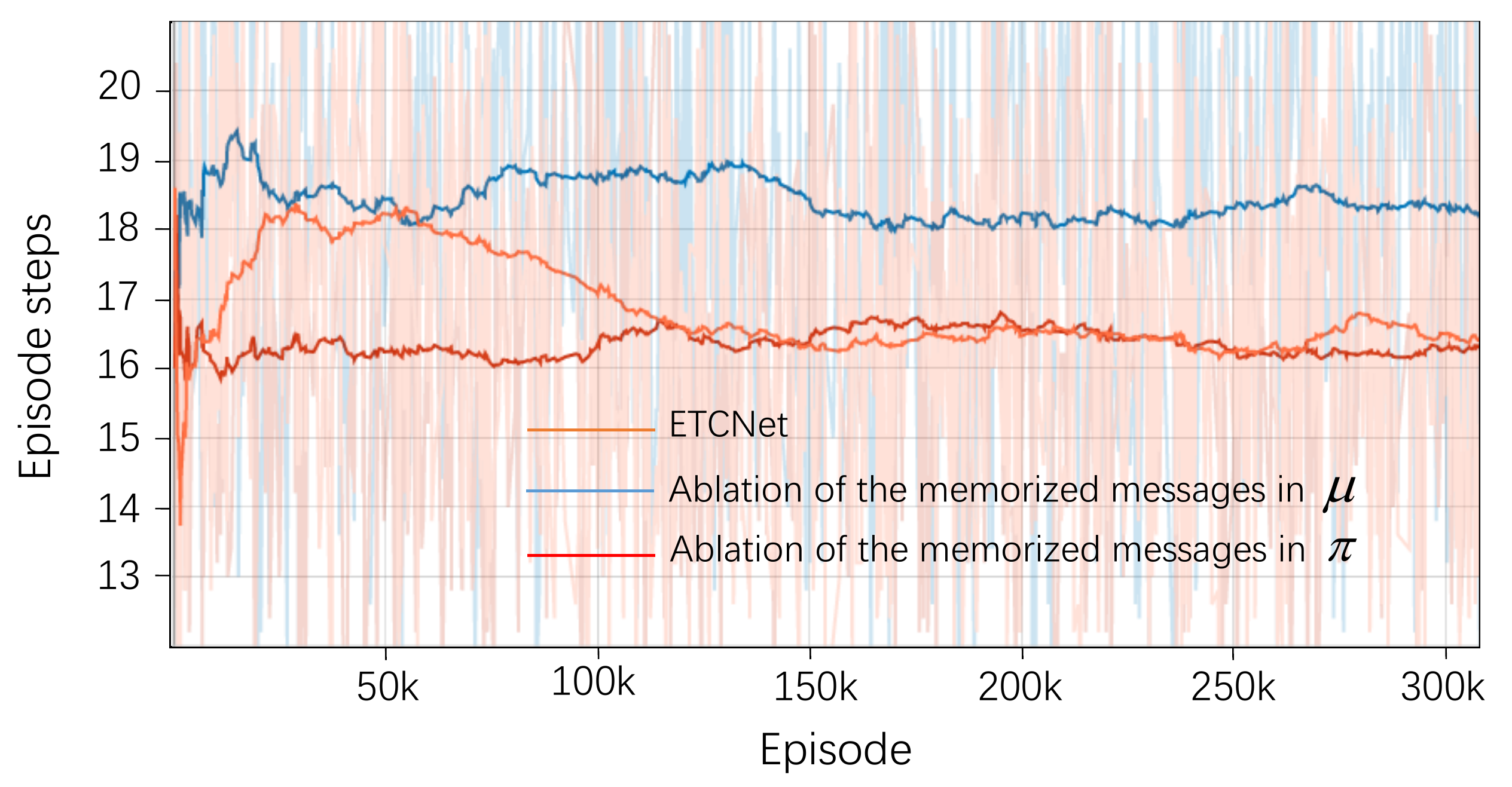}
}
\subfigure[Mean penalty per step]{
\includegraphics[width=3in, height=1.3in]{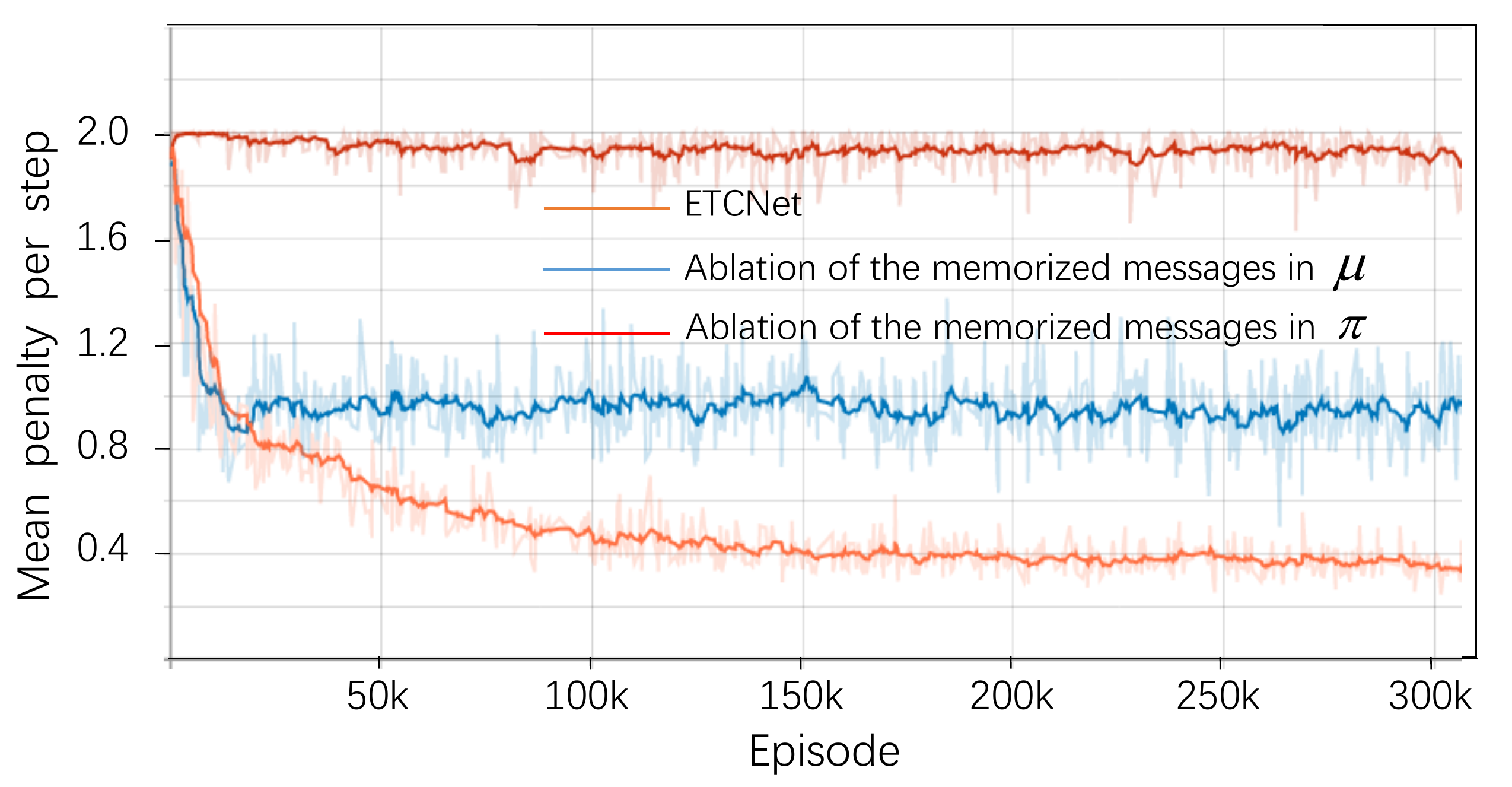}
}
\caption{Learning curves with respect to the ablation of memorized messages in $\pi_i$ and $\mu_i$.}
\label{fig:Ablation}
\end{center}
\end{figure}

To investigate the effect of the memorized messages in agent policy $\pi_i$ and gating policy $\mu_i$ , we conduct some ablation studies.
Figure \ref{fig:Ablation} shows the learning curves of ETCNet on Cooperative Navigation with different ablation.

First, we analyze the effect of the memorized messages in agent policy $\pi_i$, which is used to assist decision-making. In ablation, we pad the zero vector to ActorNet when an agent receives nothing.
We observe that the system tends to learn full communication because the failure of storing the lastly received messages greatly degrades the cooperation.

Next, we disentangle the influence of the memorized messages in gating policy $\mu_i$. The GatingNet only takes the current message as input, regardless of the lastly triggering message.
The two blue lines reveal that the ablation of the memorized messages in $\mu_i$ leads to performance degradation and bandwidth consumption.
Without knowing what has been sent in the past, the event-triggered learner has to increase sending frequency to ensure valuable messages are successfully received by others. It disturbs the learning of the multi-agent policy and deteriorates cooperation effects.

\section {Conclusion}\label{sec:Conclusion}

In this work, we propose a novel method, Event-Triggered Communication Network, for multi-agent reinforcement learning with limited-bandwidth communication.
Bandwidth limitation is mathematically transformed into a penalty threshold to restrict sending behaviours. Combined with the multi-agent optimization objective,
we establish a constrained MDP model and learns the event-triggered communication protocols. Through our experiments, we show that
ETCNet learns to send messages only when necessary while performing well in cooperation under different bandwidth constraints.

\section*{Acknowledgments}

This work was supported in part by the National Key Research and Development Program of China under Grant 2018AAA0101005 and Grant 2018AAA0102404 and in part by Huawei Noah's Ark Lab under Grant YBN2020075035.

\bibliography{ref}

\begin{thebibliography}{36}
\providecommand{\natexlab}[1]{#1}
\providecommand{\url}[1]{\texttt{#1}}
\providecommand{\urlprefix}{URL }
\expandafter\ifx\csname urlstyle\endcsname\relax
  \providecommand{\doi}[1]{doi:\discretionary{}{}{}#1}\else
  \providecommand{\doi}{doi:\discretionary{}{}{}\begingroup
  \urlstyle{rm}\Url}\fi

\bibitem[{Chu and Ye(2017)}]{parashare}
Chu, X.; and Ye, H. 2017.
\newblock Parameter sharing deep deterministic policy gradient for cooperative
  multi-agent reinforcement learning.
\newblock \emph{arXiv preprint arXiv:1710.00336} .

\bibitem[{Dimarogonas, Frazzoli, and Johansson(2012)}]{et2012}
Dimarogonas, D.~V.; Frazzoli, E.; and Johansson, K.~H. 2012.
\newblock Distributed Event-Triggered Control for Multi-Agent Systems.
\newblock \emph{IEEE Transactions on Automatic Control} 57(5): 1291--1297.

\bibitem[{Foerster et~al.(2016)Foerster, Assael, De~Freitas, and
  Whiteson}]{dial}
Foerster, J.; Assael, I.~A.; De~Freitas, N.; and Whiteson, S. 2016.
\newblock Learning to communicate with deep multi-agent reinforcement learning.
\newblock In \emph{Advances in Neural Information Processing Systems (NIPS)},
  2137--2145.

\bibitem[{Foerster et~al.(2018)Foerster, Farquhar, Afouras, Nardelli, and
  Whiteson}]{coma}
Foerster, J.~N.; Farquhar, G.; Afouras, T.; Nardelli, N.; and Whiteson, S.
  2018.
\newblock Counterfactual multi-agent policy gradients.
\newblock In \emph{Thirty-second AAAI Conference on Artificial Intelligence
  (AAAI)}, 2974--2982.

\bibitem[{Freeman(2004)}]{Freeman2004}
Freeman, R.~L. 2004.
\newblock \emph{Telecommunication System Engineering}.
\newblock Wiley.

\bibitem[{Geng et~al.(2019)Geng, Xu, Zhou, Ding, Wang, and
  Zhang}]{geng2019attention}
Geng, M.; Xu, K.; Zhou, X.; Ding, B.; Wang, H.; and Zhang, L. 2019.
\newblock Learning to cooperate via an attention-based communication neural
  network in decentralized multi-robot exploration.
\newblock \emph{Entropy} 21(3): 294.

\bibitem[{Guiasu and Shenitzer(1985)}]{principle}
Guiasu, S.; and Shenitzer, A. 1985.
\newblock The principle of maximum entropy.
\newblock \emph{The Mathematical Intelligencer} 7(1): 42--48.

\bibitem[{Hernandez-Leal et~al.(2017)Hernandez-Leal, Kaisers, Baarslag, and
  de~Cote}]{hernandez2017survey}
Hernandez-Leal, P.; Kaisers, M.; Baarslag, T.; and de~Cote, E.~M. 2017.
\newblock A survey of learning in multiagent environments: Dealing with
  non-stationarity.
\newblock \emph{arXiv preprint arXiv:1707.09183} .

\bibitem[{Jiang and Lu(2018)}]{atoc}
Jiang, J.; and Lu, Z. 2018.
\newblock Learning attentional communication for multi-agent cooperation.
\newblock In \emph{Advances in Neural Information Processing Systems (NIPS)},
  7254--7264.

\bibitem[{Kilinc and Montana(2018)}]{noisy}
Kilinc, O.; and Montana, G. 2018.
\newblock Multi-agent deep reinforcement learning with extremely noisy
  observations.
\newblock \emph{arXiv preprint arXiv:1812.00922} .

\bibitem[{Kim et~al.(2019)Kim, Moon, Hostallero, Kang, Lee, Son, and
  Yi}]{schednet}
Kim, D.; Moon, S.; Hostallero, D.; Kang, W.~J.; Lee, T.; Son, K.; and Yi, Y.
  2019.
\newblock Learning to schedule communication in multi-agent reinforcement
  learning.
\newblock \emph{arXiv preprint arXiv:1902.01554} .

\bibitem[{Kim, Cho, and Sung(2019)}]{messagedropout}
Kim, W.; Cho, M.; and Sung, Y. 2019.
\newblock Message-dropout: An efficient training method for multi-agent deep
  reinforcement learning.
\newblock In \emph{Proceedings of the AAAI Conference on Artificial
  Intelligence (AAAI)}, volume~33, 6079--6086.

\bibitem[{Levine et~al.(2016)Levine, Finn, Darrell, and Abbeel}]{levine2016end}
Levine, S.; Finn, C.; Darrell, T.; and Abbeel, P. 2016.
\newblock End-to-end training of deep visuomotor policies.
\newblock \emph{The Journal of Machine Learning Research} 17(1): 1334--1373.

\bibitem[{Lowe et~al.(2019)Lowe, Foerster, Boureau, Pineau, and
  Dauphin}]{measure}
Lowe, R.; Foerster, J.; Boureau, Y.-L.; Pineau, J.; and Dauphin, Y. 2019.
\newblock On the pitfalls of measuring emergent communication.
\newblock \emph{arXiv preprint arXiv:1903.05168} .

\bibitem[{Lowe et~al.(2017)Lowe, Wu, Tamar, Harb, Abbeel, and
  Mordatch}]{maddpg}
Lowe, R.; Wu, Y.~I.; Tamar, A.; Harb, J.; Abbeel, O.~P.; and Mordatch, I. 2017.
\newblock Multi-agent actor-critic for mixed cooperative-competitive
  environments.
\newblock In \emph{Advances in Neural Information Processing Systems (NIPS)},
  6379--6390.

\bibitem[{Mao et~al.(2017)Mao, Gong, Ni, and Xiao}]{accnet}
Mao, H.; Gong, Z.; Ni, Y.; and Xiao, Z. 2017.
\newblock ACCNet: actor-coordinator-critic cet for "learning-to-communicate"
  with deep multi-agent reinforcement learning.
\newblock \emph{arXiv preprint arXiv:1706.03235} .

\bibitem[{Mao et~al.(2019)Mao, Zhang, Xiao, Gong, and Ni}]{pruning}
Mao, H.; Zhang, Z.; Xiao, Z.; Gong, Z.; and Ni, Y. 2019.
\newblock Learning agent communication under limited bandwidth by message
  pruning.
\newblock \emph{arXiv preprint arXiv:1912.05304} .

\bibitem[{Mao et~al.(2020)Mao, Zhang, Xiao, Gong, and Ni}]{daacmp}
Mao, H.; Zhang, Z.; Xiao, Z.; Gong, Z.; and Ni, Y. 2020.
\newblock Learning multi-agent communication with double attentional deep
  reinforcement learning.
\newblock \emph{Autonomous Agents and Multi-Agent Systems} 34(1): 1--34.

\bibitem[{Peng et~al.(2017)Peng, Wen, Yang, Yuan, Tang, Long, and
  Wang}]{bicnet}
Peng, P.; Wen, Y.; Yang, Y.; Yuan, Q.; Tang, Z.; Long, H.; and Wang, J. 2017.
\newblock Multiagent bidirectionally-coordinated nets: Emergence of human-level
  coordination in learning to play starcraft combat games.
\newblock \emph{arXiv preprint arXiv:1703.10069} .

\bibitem[{Peng, Zhang, and Luo(2018)}]{peng2018attention}
Peng, Z.; Zhang, L.; and Luo, T. 2018.
\newblock Learning to communicate via supervised attentional message
  processing.
\newblock In \emph{Proceedings of the 31st International Conference on Computer
  Animation and Social Agents (CASA)}, 11--16.

\bibitem[{Pesce and Montana(2020)}]{memorydriven}
Pesce, E.; and Montana, G. 2020.
\newblock Improving coordination in small-scale multi-agent deep reinforcement
  learning through memory-driven communication.
\newblock \emph{Machine Learning} 1--21.

\bibitem[{Rashid et~al.(2018)Rashid, Samvelyan, Schroeder, Farquhar, Foerster,
  and Whiteson}]{qmix}
Rashid, T.; Samvelyan, M.; Schroeder, C.; Farquhar, G.; Foerster, J.; and
  Whiteson, S. 2018.
\newblock QMIX: monotonic value function factorisation for deep multi-agent
  reinforcement learning.
\newblock In \emph{International Conference on Machine Learning (ICML)},
  4295--4304.

\bibitem[{Shannon(1948)}]{Shannon1948}
Shannon, C.~E. 1948.
\newblock A mathematical theory of communication.
\newblock \emph{The Bell System Technical Journal} 27(3): 379--423.

\bibitem[{Shao, Zhu, and Zhao(2018)}]{shao2018starcraft}
Shao, K.; Zhu, Y.; and Zhao, D. 2018.
\newblock Starcraft micromanagement with reinforcement learning and curriculum
  transfer learning.
\newblock \emph{IEEE Transactions on Emerging Topics in Computational
  Intelligence} 3(1): 73--84.

\bibitem[{Silver et~al.(2017)Silver, Schrittwieser, Simonyan, Antonoglou,
  Huang, Guez, Hubert, Baker, Lai, Bolton et~al.}]{alphagozero}
Silver, D.; Schrittwieser, J.; Simonyan, K.; Antonoglou, I.; Huang, A.; Guez,
  A.; Hubert, T.; Baker, L.; Lai, M.; Bolton, A.; et~al. 2017.
\newblock Mastering the game of Go without human knowledge.
\newblock \emph{Nature} 550(7676): 354--359.

\bibitem[{Sim{\~o}es, Lau, and Reis(2020)}]{a3c2}
Sim{\~o}es, D.; Lau, N.; and Reis, L.~P. 2020.
\newblock Multi agent deep learning with cooperative communication.
\newblock \emph{Journal of Artificial Intelligence and Soft Computing Research}
  10(3): 189--207.

\bibitem[{Singh, Jain, and Sukhbaatar(2018)}]{ic3net}
Singh, A.; Jain, T.; and Sukhbaatar, S. 2018.
\newblock Learning when to communicate at scale in multiagent cooperative and
  competitive tasks.
\newblock \emph{arXiv preprint arXiv:1812.09755} .

\bibitem[{Sukhbaatar, Fergus et~al.(2016)}]{commnet}
Sukhbaatar, S.; Fergus, R.; et~al. 2016.
\newblock Learning multiagent communication with backpropagation.
\newblock In \emph{Advances in Neural Information Processing Systems (NIPS)},
  2244--2252.

\bibitem[{Sunehag et~al.(2017)Sunehag, Lever, Gruslys, Czarnecki, Zambaldi,
  Jaderberg, Lanctot, Sonnerat, Leibo, Tuyls et~al.}]{vdn}
Sunehag, P.; Lever, G.; Gruslys, A.; Czarnecki, W.~M.; Zambaldi, V.; Jaderberg,
  M.; Lanctot, M.; Sonnerat, N.; Leibo, J.~Z.; Tuyls, K.; et~al. 2017.
\newblock Value-decomposition networks for cooperative multi-agent learning.
\newblock \emph{arXiv preprint arXiv:1706.05296} .

\bibitem[{Szer and Charpillet(2004)}]{szer2004improving}
Szer, D.; and Charpillet, F. 2004.
\newblock Improving coordination with communication in multi-agent
  reinforcement learning.
\newblock In \emph{16th IEEE International Conference on Tools with Artificial
  Intelligence (ICTAI)}, 436--440.

\bibitem[{Vinyals et~al.(2019)Vinyals, Babuschkin, Czarnecki, Mathieu, Dudzik,
  Chung, Choi, Powell, Ewalds, Georgiev et~al.}]{alphastar}
Vinyals, O.; Babuschkin, I.; Czarnecki, W.~M.; Mathieu, M.; Dudzik, A.; Chung,
  J.; Choi, D.~H.; Powell, R.~E.; Ewalds, T.; Georgiev, P.; et~al. 2019.
\newblock Grandmaster level in StarCraft II using multi-agent reinforcement
  learning.
\newblock \emph{Nature} 575(7782): 350--354.

\bibitem[{Wang et~al.(2019)Wang, He, Yu, Qiu, An, and Rabinovich}]{imac}
Wang, R.; He, X.; Yu, R.; Qiu, W.; An, B.; and Rabinovich, Z. 2019.
\newblock Learning efficient multi-agent communication: an information
  bottleneck approach.
\newblock \emph{arXiv preprint arXiv:1911.06992} .

\bibitem[{Zhang and Lesser(2013)}]{zhang2013coordinating}
Zhang, C.; and Lesser, V. 2013.
\newblock Coordinating multi-agent reinforcement learning with limited
  communication.
\newblock In \emph{Proceedings of the 2013 international conference on
  Autonomous agents and multi-agent systems}, 1101--1108.

\bibitem[{Zhao, Chen, and Lv(2016)}]{zhao2016vehicle}
Zhao, D.; Chen, Y.; and Lv, L. 2016.
\newblock Deep reinforcement learning with visual attention for vehicle
  classification.
\newblock \emph{IEEE Transactions on Cognitive and Developmental Systems} 9(4):
  356--367.

\bibitem[{Zhou et~al.(2019)Zhou, Chen, Wen, Yang, Su, Zhang, Zhang, and
  Wang}]{fql}
Zhou, M.; Chen, Y.; Wen, Y.; Yang, Y.; Su, Y.; Zhang, W.; Zhang, D.; and Wang,
  J. 2019.
\newblock Factorized Q-learning for large-scale multi-agent systems.
\newblock In \emph{Proceedings of the First International Conference on
  Distributed Artificial Intelligence (DAI)}, 1--7.

\bibitem[{Zhu et~al.(2016)Zhu, Zhao, He, and Ji}]{zhu2016event}
Zhu, Y.; Zhao, D.; He, H.; and Ji, J. 2016.
\newblock Event-triggered optimal control for partially unknown
  constrained-input systems via adaptive dynamic programming.
\newblock \emph{IEEE Transactions on Industrial Electronics} 64(5): 4101--4109.

\end{thebibliography}

\section*{Supplement}

\subsection*{ETCNet Training Algorithm}\label{sec:Pseudocode}
The pseudocode of ETCNet training is presented in Algorithm \ref{alg:algrithm1}. First, we train the EncoderNet and ActorNet at the full communication to learn the optimal agent policy function and encoding function. After obtaining the well-trained EncoderNet and ActorNet, we keep them fixed and then train the GatingNet to get the optimal gating policy function.
After the two-stage training, ETCNet is able to perform multi-agent cooperation at a low cost of communication resources.
\begin{algorithm*}[h]
\caption{Event-Triggered Communication Network (variables and formulas are defined in the main text).}
\label{alg:algrithm1}
\LinesNumbered

Initialize the network parameters  $\lambda$, $\theta_e$, $\theta_a$, $\theta_c$, $\theta_g$, $\theta_L$, and $\theta_p$

\textbf{Training EncoderNet and ActorNet:}

Set $g_{i,t} \equiv 1$ for all $i$ and $t$

\For{ $episode = 1 \ to \ M$ }
{
    Initialize the observation $\boldsymbol{o}_t$

    \For{ $t = 1 \ to \ T$ }
    {
        $\boldsymbol{m_{t}} \gets$ calculate the message  $m_{i,t}=e_{i}\left( o_{i,t}|\theta_e \right)$ of each agent $i$

        $\boldsymbol{a_{t}} \gets$ sample the action  $a_{i,t} \sim \pi_{i}(o_{i,t},\boldsymbol{m}_{-i,t} | \theta_a)$ of each agent $i$

        Execute the actions $\boldsymbol{a_{t}}$, and
        observe the reward $\boldsymbol{r_{t}}$, next observation $\boldsymbol{o_{t+1}}$, and the approximate global state $\boldsymbol{\upsilon_{t+1}}$

        Store $(o_{i,t}, a_{i,t}, r_{i,t}, o_{i,t+1}, \upsilon_{i,t+1})$ for all agents in the replay buffer $B$

        Sample a minibatch of samples $\{(o_{i^{'},t^{'}}, a_{i^{'},t^{'}}, r_{i^{'},t^{'}}, o_{i^{'},t^{'}+1}, \upsilon_{i^{'},t^{'}+1}) \}$ from $B$

        Update $\theta_c$ by minimizing the loss (7) based on the minibatch

        Update $\theta_a$ and $\theta_e$ by minimizing the loss (8) based on the minibatch
    }
}

Obtain the optimal agent policy function $\pi^*$ and encoding function $e^*$

\textbf{Training the GatingNet:}

Set $\pi_{i}=\pi^*$ and $e_{i}=e^*$ for each agent $i$

\For{ $episode = 1 \ to \ M$ }
{
    Initialize the observation $\boldsymbol{o}_t$

    \For{ $t = 1 \ to \ T$ }
    {
        $\boldsymbol{m_{t}} \gets$ calculate the message  $m_{i,t}=e_{i}\left( o_{i,t} | \theta_e \right)$ of each agent $i$

        $\boldsymbol{g_{t}} \gets$ sample the gating action  $g_{i,t} \sim \mu_{i} ( m_{i,t}, m_{i,\hat{t}^i} | \theta_g)$ of each agent $i$

        Execute the gates $\boldsymbol{g_{t}}$, and
        update the memorized message $m_{i,\hat{t}^i}$ in gating policy $\mu_i$ and the received messages $\tilde{\boldsymbol{m}}_{-i,t}$ in agent policy $\pi_i$

        $\boldsymbol{a_{t}} \gets$ sample the action  $a_{i,t} \sim \pi_{i} \left( o_{i,t}, \tilde{\boldsymbol{m}}_{-i,t} | \theta_a \right)$ of each agent $i$

        Execute the actions $\boldsymbol{a_{t}}$, and observe the reward $\boldsymbol{r_t}$ , next observation $\boldsymbol{o_{t+1}}$, and the approximate global state $\boldsymbol{\upsilon_{t+1}}$

        Store $(o_{i,t}, m_{i,\hat{t}^i}, g_{i,t}, \tilde{\boldsymbol{m}}_{-i,t}, r_{i,t}, \lambda_t, o_{t+1}, \upsilon_{t+1})$ in the replay buffer $B$

        Sample a minibatch of samples $\{(o_{i^{'},t^{'}}, m_{i^{'},\hat{t}^{i^{'}}}, g_{i^{'},t^{'}}, \tilde{\boldsymbol{m}}_{-i^{'},t^{'}}, r_{i^{'},t^{'}}, \lambda_{t^{'}}, o_{t^{'}+1}, \upsilon_{i^{'},t^{'}+1}) \}$ from $B$

        Update  $\theta_L$ by minimize the loss (12) based on the minibatch

        Update  $\theta_g$ by minimize the loss (13) based on the minibatch

        Update  $\theta_p$ by minimize the loss (17) based on the minibatch

        Update  $\lambda_t$  according to (18) based on the minibatch

    }
}

\textbf{final} \;
\textbf{return} $\theta_e$, $\theta_a$, $\theta_g$
\end{algorithm*}

\subsection*{Environments}\label{sec:Environments}
\textbf{Cooperative Navigation.}  We adopt and modify the Cooperative Navigation task in  \cite{a3c2}. Each agent's observation includes the positions of itself, the other agent, and the other's destination (6-dimensional). The optional actions of each agent are moving up, down, left, right, and staying still (5-discrete).\\
\textbf{Predator and Prey.}  We adopt the Predator and Prey task in  \cite{a3c2}. A predator has a $5 \times 5$ size of local view, and its observation includes the position of itself and the state of its local view (25(preys in local view)+25(predators in local view)+2(self coordinates)=52-dimensional). The optional actions of predators are moving up, down, left, right, and staying still (5-discrete).

\subsection*{Baselines Modification and Configuration in Experiments}\label{sec:Configuration}
There are mainly two kinds of communication styles in existing literature. The first is broadcasting communication (e.g. A3C2 \cite{a3c2} and DIAL \cite{dial} ), in which each agent sends messages directly to all the others. The second is two-stage-point-to-point communication (e.g. SchedNet \cite{schednet} and Gated-ACML \cite{pruning}), in which the messages of all agents are first sent to a node for centralized processing, and then the processed messages are sent to all agents separately. Putting aside communication styles, the purpose of this paper is to propose a more efficient communication triggering mechanism and to show its advantages over  $\Delta Q$ (used in ATOC \cite{atoc} and Gated-ACML), top-$k$ (used in SchedNet), and random (used in Message-dropout \cite{messagedropout}) communication mechanisms. To make the comparison fair across different methods, the broadcasting communication style is adopted in our experiments.

For the implementation of different methods, common modules and hyperparameters are configured in Table \ref{tab:strucommon} and Table \ref{tab:hypercommon}, respectively. Particular modules are configured in Table \ref{tab:strudiff}. The other hyperparameters of ETCNet are presented in Table \ref{tab:hyperetc}.
We use the same network hyperparameters and Adam optimizer to update network parameters across different tasks.
There is no specific effort in fine-tuning hyperparameters and modules for better results.

\begin{table*}[h]
  \centering
  \caption{Common modules of all methods in Experiments (\# changes with the number of agents and the same below).}
    \begin{tabular}{|c|c|c|c|}
    \hline
    \hline
     ~ & Cooperative Navigation & Predator and Prey & Activation function \\
    \hline
     EncoderNet & 6-20-6 & 52-40-15 &  -relu-tanh \\
    \hline
     ActorNet & 12-40-40-5 & \#-80-40-15 &  -relu-relu-softmax \\
    \hline
     CriticNet & 12-40-40-1 & \#-120-80-1 &  -relu-relu-None \\
    \hline
    \hline
    \end{tabular}%
  \label{tab:strucommon}%
\end{table*}%
\begin{table*}[h]
  \centering
  \caption{Common hyperparameters of all methods in Experiments.}
    \begin{tabular}{|c|c|c|}
    \hline
    \hline
     ~ & Cooperative Navigation & Predator and Prey  \\
    \hline
     training time steps & 600000 & 600000  \\
    \hline
     message length & 6 & 15  \\
    \hline
     discount factor & 0.95 & 0.95  \\
    \hline
    learning rate for ActorNet/EncoderNet & 0.0002 & 0.0002  \\
    \hline
    learning rate for CriticNet & 0.0004 & 0.0004   \\
    \hline
    entropy regularization  weight & 0.01 & 0.01    \\
    \hline
    \hline
    \end{tabular}%
  \label{tab:hypercommon}%
\end{table*}%

\begin{table*}[h]
  \centering
  \caption{Particular modules of different methods in Experiments.}
    \begin{tabular}{|c|c|c|c|c|}
    \hline
    \hline
     ~ & ~ & Cooperative Navigation & Predator and Prey & Activation function \\
    \hline
     \multirow{3}*{ETCNet} & GateNet & 6-40-40-2 & 52-80-80-2 &  -relu-relu-softmax \\
    \cline{2-5}
     ~ & LagrangeNet & 12-60-60-1 & \#-120-80-1 & -relu-relu-None \\
    \cline{2-5}
     ~ & PenaltyNet & 12-60-60-1 & 52-60-40-1 & -relu-relu-relu-None \\
    \hline
      \multirow{2}*{SchedNet} & wNet & 6-20-20-1 & 52-20-20-1 &   -relu-relu-sigmoid \\
    \cline{2-5}
     ~ & QNet & 14-100-50-1 & \#-100-50-1 &   -relu-relu-None \\
    \hline
      \multirow{2}*{Gated-ACML} & QNet & 14-60-60-1 & \#-120-60-1 &   -relu-relu-None \\
    \cline{2-5}
     ~ & ClassfyNet & 6-40-40-2 & 52-80-80-2 &   -relu-relu-softmax \\
    \hline
    \hline
    \end{tabular}%
  \label{tab:strudiff}%
\end{table*}%

\begin{table*}[h]
  \centering
  \caption{Other hyperparameters of ETCNet in Experiments.}
    \begin{tabular}{|c|c|c|}
    \hline
    \hline
     ~ & Cooperative Navigation & Predator and Prey  \\\hline
    learning rate for GateNet & \multicolumn{2}{|c|}{0.0002} \\\hline
    learning rate for LagrangeNet & \multicolumn{2}{|c|}{0.0004}   \\\hline
    learning rate for PenaltyNet & \multicolumn{2}{|c|}{0.0004}  \\\hline
    \hline
    \end{tabular}%
  \label{tab:hyperetc}%
\end{table*}%

\subsection*{Demonstration of Event-Triggered Gating in Spatial Domain}\label{sec:space}
\begin{figure}[h]
   \centering
   \begin{center}
   \scriptsize
     \includegraphics*[width=2.8in]{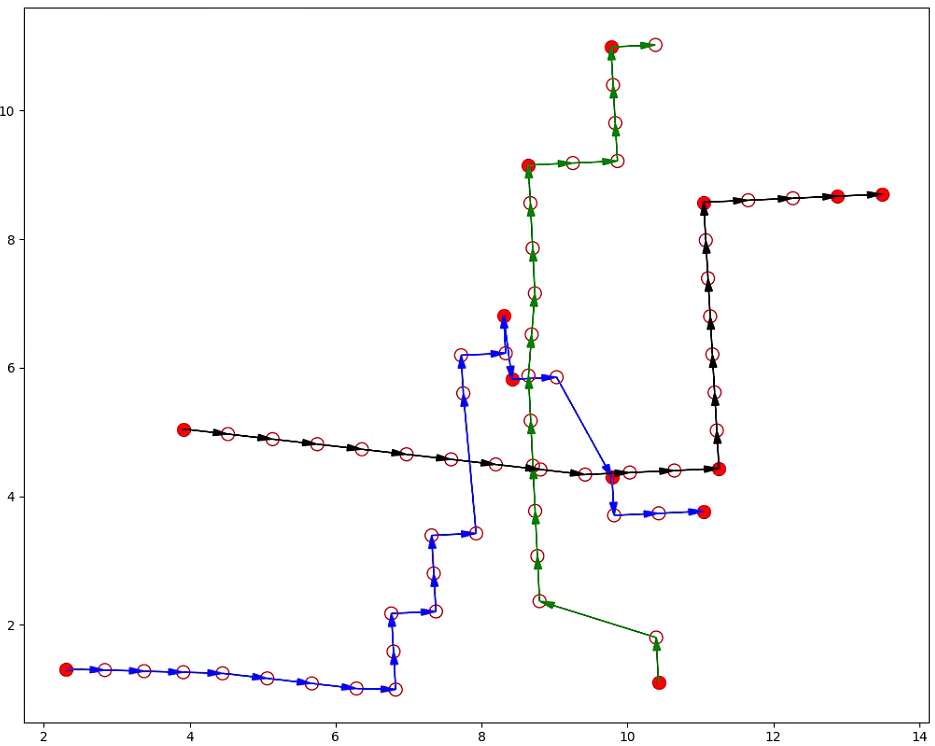}\\
   \caption{The observations  of an agent in three trajectories of Cooperative Navigation.
   The coordinates of every point are processed by PCA to compress the observation to 2D.
   The solid circle represents the agent is currently sending a message, while the hollow circle represents it is not sending a message.
   The arrows indicate the temporal order, and the arrows of  different colors represent different trajectories.}
   \label{fig:spaceng}
   \end{center}
\end{figure}

\begin{figure}[h]
   \centering
   \begin{center}
   \scriptsize
     \includegraphics*[width=2.8in]{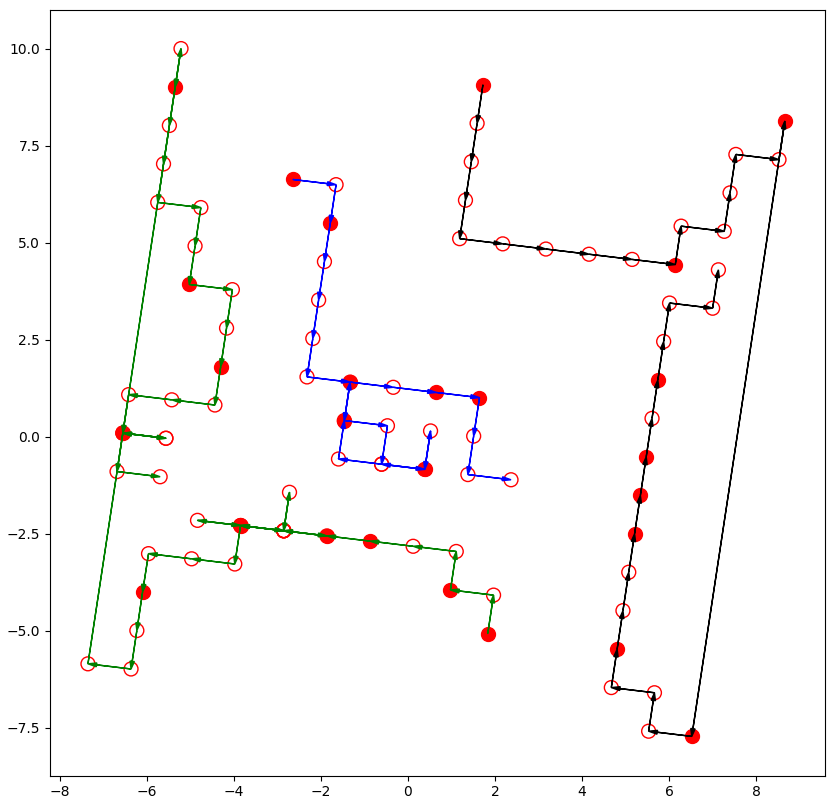}\\
   \caption{The observations  of an agent in three trajectories of Predator and Prey.    The coordinates of every point are processed by PCA to compress the observation to 2D.
   The solid circle represents the agent is currently sending a message, while the hollow circle represents it is not sending a message.
   The arrows indicate the temporal order, and the arrows of different colors represent different trajectories.}
   \label{fig:spacepp}
   \end{center}
\end{figure}

We argue that agents in ETCNet send messages not simply considering the change of observations. We demonstrate this argument by analyzing the system trajectories of two tasks obtained by ETCNet in spatial domain.
We record an agent's observations and its gating actions in three trajectories, and use principal component analysis (PCA) to compress raw observations to 2-D features. The visualization of trajectories are displayed in Figure \ref{fig:spaceng} and Figure \ref{fig:spacepp}, corresponding to Cooperative Navigation and Predator and Prey, respectively.
Some fragments show that even though there is no obvious difference in observations between two triggering moments, the ETCNet agent still decides to send messages. It reveals that the triggering condition is not simply determined by the change of observations.


\subsection*{Demonstration of Event-Triggered Gating in Time Domain}\label{sec:time}
\begin{figure*}[h]
   \centering
   \begin{center}
   \scriptsize
     \includegraphics*[width=6.5in]{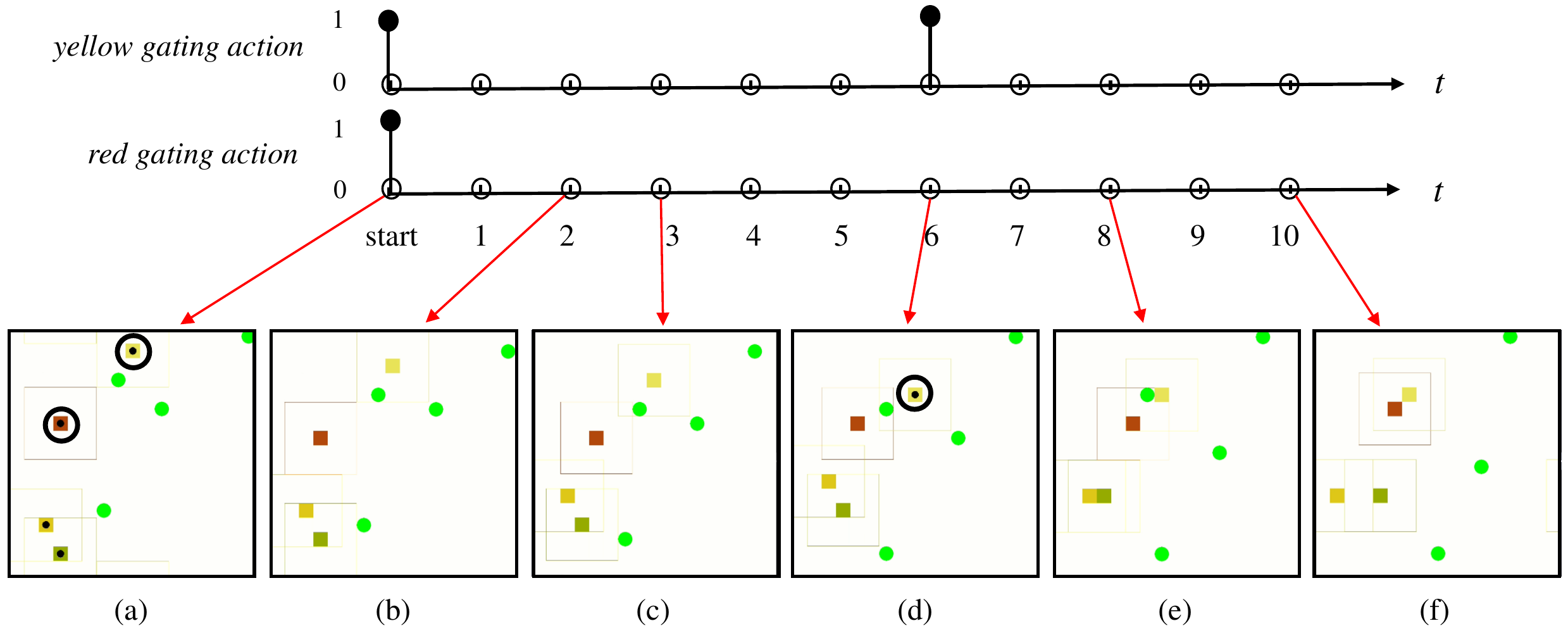}\\
   \caption{An event-triggered gating display in a fragment of a trajectory of 4-agent Predator and Prey. The square represents a predator and the green circle represents a prey. The black ring surrounding an agent indicates it is currently sending a message.}
   \label{fig:timepp}
   \end{center}
\end{figure*}

In the above Experiments, we have shown the trajectories of Cooperative Navigation obtained by ETCNet in time domain, and have argued that ETCNet agents trigger
the gating policy only when the communication is important for cooperation. The argument is also supported by the same experiment in 4-agent Predator and Prey under the desired communication probability less than $25\%$. Because the time steps are too long to elaborate, we select a representative fragment and analyze the rationality of  gating actions in
Figure \ref{fig:timepp}. We focus on the sending behaviours of the red and the yellow predators. At the starting point (a), the yellow predator sees the prey and sends a message to the others. The red predator utilizes this message to cooperate with the yellow one to surround the prey. The prey moves downwards to escape from the closest yellow predator across (b) and (c).
Even though the two predators are not communicating at these moments, the red predator still utilizes the old message and moves toward the correct direction.
At (d), the prey changes its escaping direction because of the approach of two predators. The yellow predator observes the change of prey behavior, so it sends a new message to notify the others. At and after (e), the two predators can see each other in their local views, so they stop sending messages and cooperate directly to capture the prey at (f).

\end{document}